\documentclass[dvips,reprint,a4paper,superscriptaddress,twocolumn,prb,showpacs]{revtex4}

\usepackage{amsmath}    
\usepackage{hyperref}   

\begin{document}

\title{Tomographic transform on a sphere and topological insulators}
\author{N.M. Vildanov}
\affiliation{I.E.Tamm Department of Theoretical Physics,
P.N.Lebedev Physics Institute, 119991 Moscow, Russia}


\begin{abstract}
The tomographic transform was first introduced in the field theory
literature long ago. It is closely related to Radon transform. In
this paper we show how the tomographic transform can be
implemented on a sphere and apply this result to study surface
excitations of a spherical topological insulator with a single
Dirac cone on the surface.
\end{abstract}

\maketitle

Tomographic transform finds its origin in field theory in Alan
Luther's multidimensional bosonization
construction,\cite{luther1,luther2} but explicitly it was written
down by Charles Sommerfield\cite{sommerfield} and he also coined
the term. It is also closely related to the well known Radon
transform\cite{radon} (see also the general reference on this
topic [\onlinecite{helgason}]).

Recently tomographic transform and multidimensional bosonization
based on this
transform\cite{luther1,luther2,sommerfield,aratyn1,aratyn2} were
applied in the context of topological
insulators.\cite{vildanov,moore2} The general reference on
topological insulators is [\onlinecite{review1,review2}]. Here we
will need only the fact that a strong topological insulator has an
insulating bulk and an odd number massless of Dirac fermions on
the surface. This can be viewed as a defining property of a strong
topological insulator. We will restrict the discussion to the case
of a single Dirac cone on the surface.

The aim of this paper is to implement tomographic transform on a
sphere and apply the developed formalism to study surface
excitations of a spherical topological insulator. Since the sphere
has no boundary, this is more appealing, because one does not need
to worry about boundary conditions which are simply ignored in the
case when the boundary is infinite plane. It is also obvious that
condensed matter systems are always of finite dimension.

The relation between quantum spin Hall effect and strong
topological insulators is well known and discussed in detail in
the literature. For example, in the topological band
theory\cite{fkm,moore,roy} one uses the analogy with quantum spin
Hall effect to count topological invariants of 3D insulators, and
in the topological field theory\cite{qi} both 3D topological
insulators and quantum spin Hall systems are descendants of 4D
quantum Hall effect. In this paper we show that surface
excitations of a spherical topological insulator are the sum of
edge states of quantum spin Hall disks of the same radius as the
sphere.

The spherical topological insulator was already studied in the
Refs. [\onlinecite{lee,monopole,sti}], however our approach is
completely different and it will turn out that one can still learn
some interesting facts studying this case. We also briefly discuss
a strong topological insulator having the topology of a torus and
make rigorous connection of hydrodynamic theory of surface
excitations with the topological band theory.

We discuss some technical details first. In the case of infinite
plane the tomographic transform of a function $h(x,y)$ is
\begin{equation}\label{tomographic}
    h_{\theta}(\xi)=\int_0^\infty
    {\left(\frac{k}{2\pi^3}\right)}^{1/2}dk\int \cos
    k(\xi-\xi^\prime)h(x^\prime,y^\prime)dx^\prime dy^\prime
\end{equation}
where $\xi=\hat{k}\cdot\mathbf{x}$,
$\xi^\prime=\hat{k}\cdot\mathbf{x}^\prime$,
$\hat{k}=(\cos\theta,\sin\theta)$. It satisfies the equation
\begin{equation}\label{plane}
    \int h^2(x,y)dxdy=\int_{\mathcal{R}}d\theta\int h^2_\theta(\xi)d\xi
\end{equation}
where $\mathcal{R}=\{-\pi/2\leq\theta\leq \pi/2\}$. This is
slightly modified version of the tomographic transform. We need to
find something like this on a sphere.

The starting point is the analogy with Funk transform.\cite{funk}
Funk transform of a function $f$ on a 2-sphere is defined as
\begin{equation}\label{funk}
    F f(\mathbf{x})=\int_{\mathbf{x}\in
    C(\mathbf{x})}f(\mathbf{u})ds(\mathbf{u})
\end{equation}
where $\mathbf{x}$ is a unit vector and the integration is over
the arclength $ds$ of the great circle $C(\mathbf{x})$ consisting
of all vectors perpendicular to $\mathbf{x}$. This can be viewed
as Radon transform on a sphere. In the case of the infinite plane
$C(\mathbf{x})$ are rays perpendicular to two-dimensional vector
$\mathbf{x}$.

This suggests that Euler angles could be the suitable choice as
parameter space of the tomographic representation. Indeed, there
are three angles $\alpha,\beta,\gamma$ needed to parametrize
rotations in three-dimensions and the measure in the space of
these angles is given by $d\omega=\sin\beta d\alpha d\beta
d\gamma$.\cite{landau} Two angles $\alpha$ and $\beta$ can be
chosen as the direction of $\mathbf{x}$ and the third angle
$\gamma$ as the angle denoting the position on the circle
$C(\mathbf{x})$.

Now we pose the problem. Given the function of two variables
$h(\theta,\varphi)$ we need to find a function
$h_{\alpha\beta}(\gamma)$ which is real and satisfies the equation
\begin{equation}\label{problem}
    \int_{S^2}
    h^2(\theta,\varphi)d\Omega=\int_{u}\sin\beta d\alpha d\beta \int_{0}^{2\pi}d\gamma h^2_{\alpha\beta}(\gamma)
\end{equation}
The pair $(\alpha,\beta)$ is an index, $\gamma$ is an argument,
and ${u}$ is the upper hemisphere.

To get further insight into the problem we need to study the Dirac
hamiltonian on a sphere, since multidimensional bosonization and
tomographic transform are closely related to each other. According
to the Ref.[\onlinecite{criticality}] Dirac hamiltonian on a
curved surface of a topological insulator is
\begin{equation}\label{dirac}
    H=\frac{v}{2}\left[\nabla\cdot\mathbf{n}+\mathbf{n}(\hat{p}\times\hat{\sigma})+(\hat{p}\times\hat{\sigma}\mathbf{n})\right]
\end{equation}
where $\mathbf{n}$ is a unit vector normal to the surface and
$v=1$ is the electron velocity. On a unit sphere this is
$(\hat{\sigma}\hat{l})$. There is a constant term, $\frac{1}{2}$,
which was dropped. It will be inserted in the final result.

Then one can observe that the operator $(\hat{\sigma}\hat{l})$ and
the operator $\hat{\sigma}_z\hat{l}_z$ (which is Dirac hamiltonian
on a unit circle) have the same set of eigenvalues, i.e. integers.

In \eqref{tomographic} we used Fourier analysis. Therefore we can
try to use spherical harmonic analysis in the case \eqref{problem}
and expand the function $h$ first
\begin{equation}\label{decomposition}
    h(\theta,\varphi)=\sum_l\sum_{m=-l}^{l}h_{lm}Y_{lm}(\theta,\varphi)
\end{equation}

There are a set of orthogonal functions of three Euler angles
$D^l_{m^\prime
m}(\alpha,\beta,\gamma)=e^{im^\prime\gamma}d^l_{m^\prime
m}(\beta)e^{im\alpha}$. Their explicit form can be found in
textbooks.\cite{landau} We will need only the orthogonality
relation
\begin{align}\label{ortho}
    \nonumber\int D^{j_1*}_{m_1^\prime
m_1}(\alpha,\beta,\gamma)D^{j_2}_{m_2^\prime
    m_2}(\alpha,\beta,\gamma)\frac{d\omega}{8\pi^2}\\=\frac{1}{2j_1+1}\delta_{j_1j_2}\delta_{m_1m_2}\delta_{m^\prime_1m^\prime_2}
\end{align}
and transformation properties under inversion and complex
conjugation
\begin{equation}\label{inversion}
    d^l_{m^\prime,m}(\pi-\beta)=(-1)^{l-m}d^l_{-m^\prime,m}(\beta)
\end{equation}
\begin{equation}\label{cc}
    d^{l*}_{m^\prime m}(\beta)=(-1)^{m-m^\prime}d^l_{-m^\prime,-m}(\beta)
\end{equation}
These functions are wavefunctions of a symmetric rigid rotator.
The matrix $\hat{D}^j$ gives unitary irreducible representations
of rotation group. Therefore one can try to use them to construct
the tomographic transform.

Indeed, one can check that the function
\begin{align}\label{tomtransform}
    \nonumber h_{\alpha\beta}(\gamma)=\sum_{l m}\sqrt{\frac{(2l+1)}{8\pi^2}}\left[D^l_{l
    m}(\alpha,\beta,\gamma)h_{lm}\right.\\
    \left.+D^l_{l,-m}
    (\alpha+\pi,\pi-\beta,\pi-\gamma)h_{l,-m}\right]
\end{align}
is real and satisfies Eq.\eqref{problem}. It has the symmetry
$h_{\alpha\beta}(\gamma)=h_{\pi+\alpha,\pi-\beta}(\pi-\gamma)$.
Further, if $\int h(\theta,\varphi)d\Omega=0$, i.e. $h$ describes
surface excitations of some 3D incompressible fluid, then $\int
h_{\alpha\beta}(\gamma)d\gamma=0$, which means that
$h_{\alpha\beta}(\gamma)$ can be viewed as edge excitations of
some 2D incompressible fluid.

The next step is to rewrite the Dirac hamiltonian in terms of some
spinors $\psi_{\alpha\beta}(\gamma)$. One can expand the initial
spinor $\psi=\begin{pmatrix}
  \psi_\uparrow & \psi_\downarrow \\
\end{pmatrix}^T$ as
\begin{eqnarray}
  \nonumber  \psi(\theta,\varphi)_\uparrow=\sum_{l
    m}a_{lm\uparrow}Y_{lm}(\theta,\varphi) \\
  \nonumber \psi(\theta,\varphi)_\downarrow=\sum_{l m}a_{lm\downarrow}Y_{l,m+1}(\theta,\varphi)
\end{eqnarray}
Then the hamiltonian becomes
\begin{align}
    \nonumber H=\sum_{lm}\left[m a_{lm\uparrow}^\dagger a_{lm\uparrow}-(m+1)a_{lm\downarrow}^\dagger a_{lm\downarrow}\right.\\+
    \left.\sqrt{(l-m)(l+m+1)}(a_{lm\uparrow}^\dagger
a_{lm\downarrow}+a_{lm\downarrow}^\dagger a_{lm\uparrow})\right]
\end{align}
Defining new Fermi fields
\begin{equation}\label{spinor1}
    \psi_{\alpha\beta\uparrow}(\gamma)=\sum_{lmm^\prime}a_{lm\uparrow}D^l_{m^\prime m}(\alpha,\beta,\gamma)
\end{equation}
\begin{equation}\label{spinor2}
    \psi_{\alpha\beta\downarrow}(\gamma)=\sum_{lmm^\prime}a_{lm\downarrow}D^l_{m^\prime, m+1}(\alpha,\beta,\gamma)
\end{equation}
one can write
\begin{equation}\label{}
    H=\int \psi_{\alpha\beta}^\dagger(\gamma)(\hat{\sigma}\hat{J})\psi_{\alpha\beta}(\gamma) d\omega
\end{equation}
where
\begin{equation}\label{raising}
    \nonumber J_+=J_x+iJ_y=e^{i\alpha}\left(\frac{\partial}{\partial \beta}+i\cot\beta\frac{\partial}{\partial\alpha}-i\frac{1}{\sin\beta}
    \frac{\partial}{\partial\gamma}\right)
\end{equation}
\begin{equation}\label{lowering}
    \nonumber J_-=J_x-iJ_y=e^{-i\alpha}\left(-\frac{\partial}{\partial \beta}+i\cot\beta\frac{\partial}{\partial\alpha}-i\frac{1}{\sin\beta}
    \frac{\partial}{\partial\gamma}\right)
\end{equation}
\begin{equation}\label{z}
    \nonumber \hat{J}_z=-i\frac{\partial}{\partial\alpha}
\end{equation}
This is due to relations
\begin{equation}\label{lowering1}
    \hat{J}_-D^l_{m^\prime,m+1}=\sqrt{(l-m)(l+m+1)}D^l_{m^\prime m}
\end{equation}
\begin{equation}\label{raising1}
    \hat{J}_+D^l_{m^\prime m}=\sqrt{(l-m)(l+m+1)}D^l_{m^\prime,m+1}
\end{equation}
$\hat{J}$ is the space fixed rigid rotator angular momentum
operator. We want to replace the initial problem of a Dirac
electron on a sphere with another problem: coherent states of a
symmetric rigid rotator with spin 1/2 ($\alpha$ and $\beta$ are
polar angles that determine the direction of the symmetry axis and
$\gamma$ describes rotations around this axis). This could be done
retaining only $m^\prime=\pm l,\pm(l-1)$ in the sums
\eqref{spinor1},\eqref{spinor2} and multiplying various terms with
proper coefficients. Now, when one passes to the body fixed frame,
the action of $(\hat{\sigma}\hat{J})$ on the rotated spinors
becomes diagonal (we will do this in the reverse order; see
\eqref{fermifield}). To do this one should perform Euler rotation
$U(\alpha,\beta,\gamma)$. This is easy to see in the
quasiclassical approximation when $l$ is large. In this limit the
operator $(\hat{\sigma}\hat{J})$ becomes
${(\hat{\sigma}\mathbf{n}})l$, because angular momentum is
directed along $\mathbf{n}$, and is diagonalized by rotation to a
frame in which $z$ axis points towards $\mathbf{n}$. Under
rotations of the coordinate frame, the spin part of the wave
function and its angular part transform independently. The inverse
of the operator which acts on spin part is
\begin{equation}\label{eulerrotation}
    \hat{U}(\alpha,\beta,\gamma)^{-1}=\begin{pmatrix}
      \cos\frac{\beta}{2}\cdot e^{-i(\alpha+\gamma)/2} & -\sin\frac{\beta}{2}\cdot e^{-i(\alpha-\gamma)/2} \\
      \sin\frac{\beta}{2}\cdot e^{i(\alpha-\gamma)/2} & \cos\frac{\beta}{2}\cdot e^{i(\alpha+\gamma)/2} \\
    \end{pmatrix}
\end{equation}
The transformation of angular part in the sector with momentum $l$
is described by the D-function
\begin{equation}\label{}
    \psi_{lm}=\sum_{m^\prime}D_{m^\prime m}^l\psi_{lm^\prime}
\end{equation}
There is the following formula
\begin{equation}\label{sigmaz}
    \hat{U}^{-1}\hat{\sigma}_z\hat{U}=(\mathbf{n}\hat{\sigma})
\end{equation}
where
$\mathbf{n}=(\sin\beta\cos\alpha,\sin\beta\sin\alpha,\cos\beta)$,
which will be used later.

Spinors in the rotated frame $\psi^\prime_{\alpha\beta}(\gamma)$
can be bosonized introducing boson fields
\begin{equation}\label{phi1}
    \phi^{(1)}_{\alpha\beta}(\gamma)=\sqrt{2}\sum_{l>0}\frac{e^{-\epsilon l}}{\sqrt{l}}[{b}^{(1)}_{\alpha\beta}(l)e^{il\gamma}+h.c.]
\end{equation}
\begin{equation}\label{phi2}
    \phi^{(2)}_{\alpha\beta}(\gamma)=\sqrt{2}\sum_{l>0}\frac{e^{-\epsilon l}}{\sqrt{l}}[b^{(2)\dagger}_{\alpha\beta}(l)e^{il\gamma}+h.c.]
\end{equation}
$\epsilon$ is needed for regularization of the theory and one
should take the limit $\epsilon\rightarrow 0$ in the final result.
Boson creation and annihilation operators satisfy the commutation
relations
\begin{equation}\label{comrelation}
    [b^{(i)}_{\mathbf{n}}(l),b^{(k)\dagger}_{\mathbf{n}^\prime}(j)]=\delta_{ik}\delta_{lj}\delta_{\mathbf{n},\mathbf{n}^\prime}
\end{equation}
and their time dependence is $e^{-ilt}$. Correlators of bosonic
exponents are
\begin{align}\label{correlator}
    \nonumber \langle\exp[-{\phi^{(1)}_{\mathbf{n}}(t,\gamma)}]\exp[{\phi^{(1)}_{\mathbf{n^\prime}}(0,\gamma^\prime)}]\rangle\\=
    \nonumber \frac{\epsilon^2}{\left[1-e^{-i(t-\gamma+\gamma^\prime)-\epsilon}\right]^2}\delta_{\mathbf{n},\mathbf{n}^\prime}\\=
    \delta_{\mathbf{n},\mathbf{n}^\prime}\cdot\epsilon^2\sum_{l=0}^\infty (l+1)e^{-il(t-\gamma+\gamma^\prime)-\epsilon l}
\end{align}
and for $\phi^{(2)}$ one should change the sign of $\gamma$ and
$\gamma^\prime$.

The usual Fermi field is given by
\begin{flalign}\label{fermifield}
   \nonumber\hat\psi(\theta,\varphi)=\frac{1}{\sqrt{2\pi}\epsilon}\sum_{lm}\int_ud\Omega\int_{0}^{2\pi}\frac{d\gamma}{2\pi}~(-1)^mY_{lm}(\theta,\varphi)\hat{U}^{-1} &\\
    \times\left\{e^{-i{(t-\gamma)}/{2}}
    e^{\phi^{(1)}}\left[\begin{pmatrix}
      D^l_{-l,-m} \\
      0 \\
    \end{pmatrix}+\frac{\sqrt{\frac{l}{l+1}}e^{it}}{\sqrt{2l+1}}\begin{pmatrix}
      \sqrt{2l}D^l_{l,-m} \\
      D^l_{l-1,-m} \\
    \end{pmatrix}\right] \right.\nonumber\\
    \left. {}+e^{-i{(t+\gamma)}/{2}}
    e^{\phi^{(2)}}\left[\begin{pmatrix}
      0 \\
      D^l_{l,-m} \\
    \end{pmatrix}+\frac{\sqrt{\frac{l}{l+1}}e^{it}}{\sqrt{2l+1}}\begin{pmatrix}
      D^l_{-l+1,-m} \\
      \sqrt{2l}D^l_{-l,-m} \\
    \end{pmatrix}\right]\right\}
\end{flalign}
Here $\hat{U}$ stands for $\hat{U}(\alpha,\beta,\gamma)$, the
arguments of the D-functions are $\alpha,\beta,\gamma$,
$\phi^{(i)}$ is the shortening for
$\phi^{(i)}_{\alpha\beta}(\gamma)$ and $d\Omega=\sin\beta d\alpha
d\beta$. This expression seems complicated, but it has a clear
meaning. We will demonstrate this by deriving equations of motion
for this field. First, we compute the derivative $id/dt$. Due to
simple equations of motions for $\phi^{(1)}$ and $\phi^{(2)}$
\begin{equation}\label{eqofmotion}
    \partial_t\phi^{(1)}+\partial_\gamma\phi^{(1)}=0,\quad
    \partial_t\phi^{(2)}-\partial_\gamma\phi^{(2)}=0
\end{equation}
one can replace this derivative by $\mp id/d\gamma$ and then
integrate by parts. $d/d\gamma$ now is multiplication by a
constant. On the other hand, action of $(\hat{\sigma}\hat{l})$ on
rhs of \eqref{fermifield} can be replaced by
$-(\hat{\sigma}\hat{J})$, $\hat{J}$ acting on D-functions only,
not on $\hat{U}$ or bosonic exponents (this is due to formulas
\eqref{lowering1} and \eqref{raising1}). Using
$(\hat{\sigma}\hat{J})\hat{U}^{-1}=\hat{U}^{-1}(\hat{\sigma}\hat{P})$
($\hat{J}$ does not act on $\hat{U}$) where $\hat{P}$ is the body
fixed rigid rotator angular momentum operator
\begin{equation}
    \nonumber \hat{P}_{+}=\hat{P}_1-i\hat{P}_2=e^{i\gamma}\left(-\frac{\partial}{\partial\beta}-i\cot\beta\frac{\partial}{\partial\gamma}+
    i\frac{1}{\sin\beta}\frac{\partial}{\partial\alpha}\right)
\end{equation}
\begin{equation}
    \nonumber \hat{P}_{-}=\hat{P}_1+i\hat{P}_2=e^{-i\gamma}\left(\frac{\partial}{\partial\beta}-i\cot\beta\frac{\partial}{\partial\gamma}+
    i\frac{1}{\sin\beta}\frac{\partial}{\partial\alpha}\right)
\end{equation}
\begin{equation}\label{}
    \nonumber \hat{P}_z=-i\frac{\partial}{\partial\gamma}
\end{equation}
which has anomalous commutation relations
$[\hat{P}_i,\hat{P}_j]=-i\varepsilon_{ijk}\hat{P}_k$ and acts on
the D-functions according to
\begin{equation}\label{lowering2}
    \nonumber \hat{P}_-D^l_{m^\prime+1,m}=\sqrt{(l-m^\prime)(l+m^\prime+1)}D^l_{m^\prime, m}
\end{equation}
\begin{equation}\label{raising2}
    \nonumber \hat{P}_+D^l_{m^\prime m}=\sqrt{(l-m^\prime)(l+m^\prime+1)}D^l_{m^\prime+1,m}
\end{equation}
\begin{equation}\label{zcomponent}
    \nonumber \hat{P}_zD^l_{m^\prime m}=m^\prime D^l_{m^\prime m}
\end{equation}
and the fact that spinors in square brackets are eigenfunctions of
$(\hat{\sigma}\hat{P})$ with eigenvalues $-l$ and $l+1$, one can
show that
\begin{equation}
    \nonumber
    i\frac{d\psi}{dt}=\left[\frac{1}{2}+(\hat{\sigma}\hat{l})\right]\psi
\end{equation}
This is correct equation of motion.

Now we will outline the derivation pairwise correlation functions
$\langle\psi\psi^\dagger\rangle$ and
$\langle\psi^\dagger\psi\rangle$ from \eqref{fermifield} (details
of the calculations can be found in appendix \ref{appendix}).
Using \eqref{correlator} one can see that integration over
$\gamma$ and $\gamma^\prime$ in the fermionic correlator picks up
the sector with momentum $l$ from the bosonic correlator. First of
the two spinors in the square brackets in \eqref{fermifield} are
important for the calculation of the correlator
$\langle\psi\psi^\dagger\rangle$, and the second ones for the
correlator $\langle\psi^\dagger\psi\rangle$ only. There are two
independent contributions to the fermionic correlator emerging
from two different bosonic exponents. The part obtained from the
terms in the second quare brackets in \eqref{fermifield} can be
converted to the the part obtained from the first square brackets,
but where the integration is over the lower hemisphere, by using
transformation properties of D-functions
\eqref{inversion},\eqref{cc}. Calculating the emerging integrals
of the product of three D-functions
($\cos\beta=d^1_{00}(\beta)$,~$\sin\beta=\sqrt{2}d^1_{10}(\beta)$)
which are now over the entire sphere, using formulas in
[\onlinecite{landau}], it can be shown that one obtains correct
expressions for fermionic correlators. For example
\begin{align}\label{green}
    \nonumber \langle\psi_\uparrow(t,\theta,\varphi)\psi^\dagger_\uparrow(0,\theta^\prime,\varphi^\prime)\rangle \qquad\qquad\qquad\qquad\qquad\\
    \nonumber=\sum_{lm}e^{-it(l+1/2)}\frac{l+m+1}{2l+1}
    Y_{lm}(\theta,\varphi)Y_{lm}^*(\theta^\prime,\varphi^\prime)
\end{align}
which coincides with the expression obtained by conventional means
using eigenfunctions of the operator $(\hat{\sigma}\hat{l})$ with
eigenvalue $l$
\begin{equation}\label{l}
    \sqrt{\frac{l+m+1}{2l+1}}\begin{pmatrix}
      1 \\
      0 \\
    \end{pmatrix}Y_{lm}+\sqrt{\frac{l-m}{2l+1}}\begin{pmatrix}
      0 \\
      1 \\
    \end{pmatrix}Y_{l,m+1}
\end{equation}
and eigenvalue $-l-1$
\begin{equation}\label{l2}
    -\sqrt{\frac{l-m}{2l+1}}\begin{pmatrix}
      1 \\
      0 \\
    \end{pmatrix}Y_{lm}+\sqrt{\frac{l+m+1}{2l+1}}\begin{pmatrix}
      0 \\
      1 \\
    \end{pmatrix}Y_{l,m+1}
\end{equation}
The necessity of introducing additional factors $\sqrt{l/(l+1)}$
in Eq.\eqref{fermifield} can be seen by explicit calculations (see
appendix \ref{appendix}).

The operators \eqref{fermifield} do not anticommute. Therefore one
should insert Klein factors in \eqref{fermifield} that ensure
anticommutation relations. In the continuum case\cite{luther1} one
uses the limit $k\rightarrow 0$ of some boson creation operators
for this purpose, where $k$ is momentum. This is not possible in
the discrete case we are dealing with. We will introduce some
finite but very small compressibility of the liquid and static
bosonic field $b_{0\alpha\beta}^{(i)}$ instead. Then if one adds a
term $b^{(i)}_{0\alpha\beta}/\delta^{1/2}$ to
$\phi^{(i)}_{\alpha\beta}$ and inserts the operator
\begin{equation}\label{klein}
    \nonumber \hat{O}_\Omega=e^{{\pi i}
    \sum_{\Omega^\prime\leq\Omega}[b_{0\Omega}^{(1)}+b_{0\Omega}^{(2)}]\delta^{1/2}}
\end{equation}
into the integrand in \eqref{fermifield}, then \eqref{fermifield}
will satisfy the required anticommutation relations
\begin{equation}\label{anticommutation}
    \nonumber
    \{\psi_i^\dagger{(t,\mathbf{x})},\psi_k{(t,\mathbf{x}^\prime)}\}=\delta_{ik}\delta(\mathbf{x}-\mathbf{x}^\prime)
\end{equation}
We assume that $\delta$ is small. Introducing the operator
$b_{0\alpha\beta}^{(i)}$ into the theory modifies the electronic
density operator: there will be an extra term proportional to
$\delta^{1/2}$ which tends to 0 in the limit $\delta\rightarrow
0$. This is consistent with the incompressibility of the liquid.

One can trace the analogy between the scheme developed in this
paper and that of Luther's almost in every step. Luther also
showed that bosonic exponents in his scheme lead to correct
expressions for general correlation functions. His analysis relies
only on the two-point correlation functions, equations of motion
for the fermi fields and anticommutation relations. One can
directly extend this analysis to the present case and there is no
need to repeat them here.

The formula \eqref{fermifield} has no useful applications. It was
derived only to show that bosonization on a sphere can be
consistently carried out.

Though we mentioned the relation of our analysis to topological
insulators throughout the paper several times, it was quite
abstract. Now we will make this relation more concrete. Following
the Ref.[\onlinecite{vildanov}] one can describe low lying
excitations of a strong topological insulator as surface
deformations of a two-component 3D incompressible liquid confined
by a smooth potential well with the Hamiltonian
\begin{equation}\label{energy}
    {H}=\frac{1}{2}\rho_0eER^2\int
    \left[h^2_1(\theta,\varphi)+h^2_2(\theta,\varphi)\right]d\Omega
\end{equation}
where $E$ is the electric field of the confining potential on the
surface, $\rho_0$ is the density of the 3D electronic liquid,
$\int h_i(\theta,\varphi) d\Omega=0$, $i=1,2$ and $R$ is the
radius of the sphere.  This theory is a modification of the
hydrodynamic theory of the edge excitations of fractional quantum
Hall systems.\cite{wen} Using \eqref{problem} and
\eqref{tomtransform} one can rewrite \eqref{energy} as follows
\begin{equation}\label{energy2}
    H=\frac{1}{2}neER\int_u d\Omega\int_0^{2\pi}
    \left[h^2_{1\alpha\beta}(\gamma)+h^2_{2\alpha\beta}(\gamma)\right]d\gamma
\end{equation}
where $h_{i\alpha\beta}$ can be viewed as surface deformations of
a 2D incompressible fluid with the density $n=C\rho_0R$ and having
a shape of a disk of radius $R$, $C$ is a numeric constant
($h_{\alpha\beta}$ has been rescaled by a factor $C^{-1/2}$
compared to \eqref{tomtransform}). We have assumed in
\eqref{energy2} that electric field of the potential which
confines this liquid is $E$. In analogy with the paper
[\onlinecite{vildanov}] we assume that $h_{i\alpha\beta}(\gamma)$
are edge states of quantum spin Hall system, i.e. they have
equations of motion
\begin{equation}\label{eqofmotion2}
    \partial_th_{i\alpha\beta}(\gamma)=(-1)^iv\partial_\gamma h_{i\alpha\beta}(\gamma)
\end{equation}
where $v=eE/hn$. Thus the constant $C$ is not arbitrary:
$C=eE/h\rho_0vR$. The theory \eqref{energy2}, \eqref{eqofmotion2}
can be easily quantized
\begin{equation}\label{quantized}
    H=\int_ud\Omega \sum_{l>0}\sum_{i=1,2}vlb^{(i)\dagger}_{\alpha\beta}(l)b^{(i)}_{\alpha\beta}(l)
\end{equation}
where the boson operators $b^{(i)}_{\alpha\beta}$ were introduced
earlier (\ref{phi1}-\ref{comrelation}).

Thus we have found that surface excitations of a spherical
topological insulator with a single Dirac cone on the surface are
the sum of quantum spin Hall edge states. Since the transform
\eqref{tomtransform} is non-local, every edge state that has a
fixed location in the tomographic representation is spread over
the entire surface of the topological insulator.

Now we will briefly discuss a toric topological insulator for
which one can make a direct connection with the topological band
theory. This case is simpler than a spherical case.

Suppose that there is an ideal crystal of a strong topological
insulator with atomic spacing $a$ having $N\gg 1$ unit cells along
$x$ and $y$ directions and periodic boundary conditions imposed in
these directions. $z=0$ is one of the surfaces of this crystal.
The function $h(x,y)$ defined on this surface and having the
property $\int h(x,y)dxdy=0$ can be expanded as $(L=Na)$
\begin{equation}\label{decomposition2}
    h(x,y)=\sum_{mn}h_{mn}e^{2\pi i(mx+ny)/L}
\end{equation}
and there can be defined a new function
\begin{align}\label{newfunc}
    \nonumber h_{(mn)}(\xi)=({m^2+n^2})^{1/4}\sum_{l=1}^\infty
    [h_{lm,ln}e^{2\pi il\xi/L_{(mn)}}\\
    +h_{-lm,-ln}e^{-2\pi il\xi/L_{(mn)}}]
\end{align}
where
\begin{equation}\label{period}
    \nonumber L_{(mn)}=Na_{(mn)},\quad a_{(mn)}=\frac{a}{\sqrt{m^2+n^2}}
\end{equation}
and $(mn)$ denotes a pair of coprime integers $m$ and $n$ such
that $m>0$ and $n$ arbitrary or one of the pairs $m=0, n=1$ or
$m=1, n=0$. $(mn)$ are in fact Miller indices and $l$ has the
meaning of a winding number. It is easy to see now that
\begin{equation}\label{tomographic3}
    \int h^2(x,y)dx
    dy=L\sum_{(mn)}\int_0^{L_{(mn)}}h^2_{(mn)}(\xi)d\xi
\end{equation}
where the sum is over all possible pairs $(mn)$. This means that
we have rewritten the 2D hamiltonian as a sum of 1D hamiltonians
over closed loops of length $L_{(mn)}$.

One can see that if $h(x,y)$ describes surface excitations of 3D
incompressible fluid, $h_{00}=0$, then $h_{(mn)}(\xi)$ describes
edge excitations of a 2D incompressible fluid, $\int
h_{(mn)}(\xi)d\xi=0$.

The above procedure applies also to 2D fermionic hamiltonian
\begin{equation}\label{fermtom}
    \frac{2\pi}{L}\sum_{l,(mn)}\hat{\psi}^\dagger_{lm,ln}vl(m\hat{\sigma}_x+n\hat{\sigma}_y)\hat{\psi}_{lm,ln}
\end{equation}
The operator $m\hat{\sigma}_x+n\hat{\sigma}_y$ can be diagonalized
by a rotation. Diagonalized hamiltonian is
$2\pi\hat{\sigma}_zv\sqrt{m^2+n^2}/L$. Eq. \eqref{fermtom}
describes surface excitations of a strong topological insulator
for small momenta $l\sqrt{m^2+n^2}\ll N$. Surface deformations of
two incompressible liquids $h_{i(mn)}(t,\xi)$, $i=1,2$ will
satisfy equations of motion
\begin{equation}\label{}
    \nonumber \partial_th_{i(mn)}(t,\xi)=(-1)^i\partial_\xi h_{i(mn)}(t,\xi)
\end{equation}
These equations are consistent with the spectrum $2\pi
vl\sqrt{m^2+n^2}/L$ of the fermionic hamiltonian \eqref{fermtom}
and the exponent $e^{\pm 2\pi il\xi/L_{(mn)}}$ we have chosen in
\eqref{newfunc}. It is clear that one can construct bosonization
scheme in analogy with the Ref. [\onlinecite{luther1}].

Now it is easy to make parallel with the $Z_2$ invariants of 2D
band structures.\cite{km} Let $\hat{H}(k_x,k_y,k_z)$ be a tight
binding Hamiltonian of a strong topological insulator (see, e.g.
the Ref. \onlinecite{review2}) and $\hat{H}_{(mn)}(2\pi
l/L_{(mn)},k_z)=\hat{H}(2\pi lm/L,2\pi ln/L,k_z)$.  There is one
to one correspondence between $\hat{H}_{(mn)}(2\pi l/L_{(mn)},z)$
and the edge state hamiltonians \eqref{fermtom} for fixed
direction $(mn)$. The hamiltonian $\hat{H}_{(mn)}(2\pi
l/L_{(mn)},k_z)$ is periodic with the period
$\frac{2\pi}{a_{(mn)}}$ along $k_\xi$ direction. This means that
the 2D lattice has Brillouin zone
$\left\{-\frac{\pi}{a_{(mn)}}\leq k_\xi\leq
\frac{\pi}{a_{(mn)}},-\frac{\pi}{a}\leq k_z \leq
\frac{\pi}{a}\right\}$. It follows from the topological band
theory that all hamiltonians $\hat{H}_{(mn)}$ for fixed $(mn)$ are
hamiltonians supporting quantum spin Hall effect. The easiest way
to see this is to use the simple counting argument due to
Roy\cite{roy2}, or the approach of the papers
[\onlinecite{fkm,inversion_symmetry}]. In other words $Z_2$
invariants of the planes $(mn)$ are all equal to 1 in the case of
a strong topological insulator. This means that the fields
$h_{i(mn)}$, $i=1,2$ (in the hydrodynamic theory) are the edge
states of quantum spin Hall systems.

There were some difficulties in the continuous
case.\cite{vildanov} It is clear now that these difficulties can
be overcome by considering a discrete version of the tomographic
transform.

The assumption of equal number of sites along $x$ and $y$
directions is not essential. If there are different number of
sites, then this would only lead to unnecessary complication of
the analysis.

In summary, we have shown how the tomographic transform can be
implemented on a sphere.  Using this transform, we constructed
Luther's version of the bosonization on a sphere and showed that
surface excitations of a spherical topological insulator with a
single Dirac cone on the surface are the sum of quantum spin Hall
edge states. However, this is not true for the entire topological
insulator: strong topological insulator can not be presented as a
sum of two-dimensional topological insulators. We also constructed
discrete version of the tomographic transform on a torus and
rigorously showed how the connection can be made between
hydrodynamic theory of surface excitations of a non-interacting
topological insulator and the topological band theory.

\begin{acknowledgements}
    I would like to acknowledge S.M. Apenko, V.V. Losyakov, and
    A.G. Semenov for interesting discussions.
\end{acknowledgements}

\appendix

\section{Calculation of correlation functions}\label{appendix}

\begin{widetext}
In this appendix, we will show how to compute two-point
correlation functions starting from the Eq. \eqref{fermifield}.
After calculation of bosonic exponents and integration over
$\gamma$ and $\gamma^\prime$ one obtains
\begin{flalign*}\label{}
    \nonumber \langle\hat{\psi}(t,\theta,\varphi)\hat{\psi}^\dagger(0,\theta^\prime,\varphi^\prime)\rangle=
    \frac{1}{2\pi}\sum_{lm_1m_2}(l+1)e^{-it(l+1/2)}(-1)^{m_1-m_2}Y_{lm_1}(\theta,\varphi)Y_{lm_2}^*(\theta^\prime,\varphi^\prime)\\
    \times\left[\int_ud\Omega\frac{1+\hat{\sigma}\mathbf{n}}{2}
    D^l_{-l,-m_1}(\alpha,\beta,0)D^{l*}_{-l,-m_2}(\alpha,\beta,0)+
    \int_ud\Omega\frac{1-\hat{\sigma}\mathbf{n}}{2}
    D^l_{l,-m_1}(\alpha,\beta,0)D^{l*}_{l,-m_2}(\alpha,\beta,0)\right]
\end{flalign*}
Using Eqs.\eqref{inversion} and \eqref{cc} it can be shown that
the second integral in square brackets equals
$$
    \int_{L}d\Omega\frac{1+\hat{\sigma}\mathbf{n}}{2}
    D^l_{-l,-m_1}(\alpha,\beta,0)D^{l*}_{-l,-m_2}(\alpha,\beta,0)
$$
where `$L$' means that integration is over the lower hemisphere.
The integral
$$
    \int d\Omega\frac{1+\hat{\sigma}\mathbf{n}}{2}
    D^l_{-l,-m_1}(\alpha,\beta,0)D^{l*}_{-l,-m_2}(\alpha,\beta,0)
$$
can be calculated using the following integral of the product of
three D-functions
\begin{equation}\label{dintegral}
    \int
    D^{j_1}_{m_1^\prime m_1}(\omega)D^{j_2}_{m_2^\prime m_2}(\omega)D^{j_3}_{m_3^\prime
    m_3}(\omega)\frac{d\omega}{8\pi^2}=\begin{pmatrix}
      j_1 & j_2 & j_3 \\
      m_1^\prime & m_2^\prime & m_3^\prime \\
    \end{pmatrix}
    \begin{pmatrix}
      j_1 & j_2 & j_3 \\
      m_1 & m_2 & m_3 \\
    \end{pmatrix}
\end{equation}
and the fact that
\begin{equation*}\label{}
    \frac{1\pm\cos\beta}{2}=d^1_{1,\pm
    1}(\beta),\qquad\frac{1}{\sqrt{2}}\sin\beta=d^1_{10}(\beta)
\end{equation*}
Here $\begin{pmatrix}
      j_1 & j_2 & j_3 \\
      m_1 & m_2 & m_3 \\
    \end{pmatrix}$ is a $3j$-symbol which is non-zero only when $m_3=-m_1-m_2$. For $j_1=j_2$ and $j_3=1$,
$3j$-symbols can be easily calculated using tables in
\cite{landau}. For example,
$$
    \begin{pmatrix}
      l & l & 1 \\
      m & -m & 0 \\
    \end{pmatrix}=(-1)^{l-m}\frac{2m}{[2l(2l+1)(2l+2)]^{1/2}},\qquad \begin{pmatrix}
      l & l & 1 \\
      m & -m-1 & 1 \\
    \end{pmatrix}=(-1)^{l-m}{\left[\frac{2(l-m)(l+m+1)}{2l(2l+1)(2l+2)}\right]^{1/2}}
$$
The result coincides with the one obtained by using the
eigenfunction \eqref{l} of the operator $(\hat{\sigma}\hat{l})$.

For the other correlator $\langle\psi^\dagger\psi\rangle$, after
integration over $\gamma$ and $\gamma^\prime$, one has
\begin{align}\label{correlator2}
    \langle\hat{\psi}^\dagger(0,\theta^\prime,\varphi^\prime)\hat{\psi}(t,\theta,\varphi)\rangle=
    \frac{1}{2\pi}\sum_{lm_1m_2}\frac{l}{2l+1}e^{it(l+1/2)}(-1)^{m_1-m_2}Y_{lm_1}(\theta,\varphi)Y_{lm_2}^*(\theta^\prime,\varphi^\prime)\\
    \times \int_ud\Omega \nonumber
    \left\{
    \left[2lD^{l}_{l,-m_1}(\alpha,\beta,0)D^{l*}_{l,-m_2}(\alpha,\beta,0)\frac{1+\hat{\sigma}\mathbf{n}}{2}+
    D^{l}_{l-1,-m_1}(\alpha,\beta,0)D^{l*}_{l-1,-m_2}(\alpha,\beta,0)
    \frac{1-\hat{\sigma}\mathbf{n}}{2}\right.\right.\nonumber \\
    \left.\left.+\hat{\sigma}_+(\alpha,\beta)\sqrt{2l}D^{l}_{l,-m_1}(\alpha,\beta,0)D^{l*}_{l-1,-m_2}(\alpha,\beta,0)+
    \hat{\sigma}_{-}(\alpha,\beta)\sqrt{2l}D^{l}_{l-1,-m_1}(\alpha,\beta,0)D^{l*}_{l,-m_2}(\alpha,\beta,0)
    \right]\right.\nonumber\\
    \left.+
    \left[D^{l}_{-l+1,-m_1}(\alpha,\beta,0)D^{l*}_{-l+1,-m_2}(\alpha,\beta,0)\frac{1+\hat{\sigma}\mathbf{n}}{2}+
    2lD^{l}_{-l,-m_1}(\alpha,\beta,0)D^{l*}_{-l,-m_2}(\alpha,\beta,0)
    \frac{1-\hat{\sigma}\mathbf{n}}{2}\right.\right.\nonumber \\
    \left.\left.+\hat{\sigma}_+(\alpha,\beta)\sqrt{2l}D^{l}_{-l+1,-m_1}(\alpha,\beta,0)D^{l*}_{-l,-m_2}(\alpha,\beta,0)+
    \hat{\sigma}_{-}(\alpha,\beta)\sqrt{2l}D^{l}_{-l,-m_1}(\alpha,\beta,0)D^{l*}_{-l+1,-m_2}(\alpha,\beta,0)\phantom{\frac{1}{2}}\right]
    \right\}
\end{align}
where
$$
    \hat{\sigma}_\pm(\alpha,\beta)=\frac{1}{2}\hat{U}^{-1}(\alpha,\beta,0)\hat{\sigma}_\pm\hat{U}(\alpha,\beta,0)
$$
or explicitly
$$
    \hat{\sigma}_+(\alpha,\beta)=\hat{\sigma}_{-}^T(-\alpha,\beta)=\frac{1}{2}\begin{pmatrix}
      \sin\beta & e^{-i\alpha}(1+\cos\beta) \\
      e^{i\alpha}(1-\cos\beta) & \sin\beta \\
    \end{pmatrix}
$$
Let $\hat{A}$ be the expression in the first square brackets in
\eqref{correlator2} and $\hat{B}$ in the second. Then using the
Eqs. \eqref{inversion} and \eqref{cc} it can be shown that
$$
    \int_L\hat{A}d\Omega=\int_u\hat{B}d\Omega
$$
The integral $\int \hat{A}d\Omega$ can be calculated using
\eqref{dintegral}. For example,
$$
    \int A_{11}d\Omega=\frac{l-m_1}{l}\delta_{m_1m_2}
$$
and
$$
    \langle\psi^\dagger_\uparrow(0,\theta^\prime,\varphi^\prime)\psi_\uparrow(t,\theta,\varphi)\rangle=
    \sum_{lm}\frac{l-m}{2l+1}e^{it(l+1/2)}Y_{lm}(\theta,\varphi)Y_{lm}^*(\theta^\prime,\varphi^\prime)
$$
which coincides with the expression obtained from \eqref{l2} by
conventional means.

\end{widetext}

\end{document}